\documentclass[12pt,a4paper]{article}

\usepackage[english]{babel}
\usepackage[T2A]{fontenc}
\usepackage[cp1251]{inputenc}
\usepackage{amsmath}
\usepackage{graphicx}
\usepackage{amssymb}
\usepackage{color}
\usepackage{amsfonts}
\usepackage{wrapfig}
\usepackage{caption}

\usepackage{hyperref}
\usepackage{physics}

\begin{document}
	
	\binoppenalty=10000
	\relpenalty=10000
	
\begin{center}
	\textbf{\Large{Supercritical Crossover Lines in the Cell Fluid Model}}
\end{center}

\vspace{0.3cm}

\begin{center}
 O.A.~Dobush\footnote{e-mail:  dobush@icmp.lviv.ua}, M.P.~Kozlovskii, R.V.~Romanik
\end{center}

\begin{center}
	Institute for Condensed Matter Physics of the National Academy of Sciences of
	Ukraine \\ 1, Svientsitskii Street, 79011 Lviv, Ukraine
\end{center}

	 \vspace{0.2cm}

	\begin{center}
 \small \textbf{Abstract} 
	\end{center}
	
	\small A cell fluid model with a modified Morse potential is studied. The supercritical states are considered with respect to a possibility to build a separation boundary between liquid-like and gas-like bahaviors. Three different lines are calculated that can be used for this purpose: the locus of the isothermal compressibility maxima, the locus of the thermal expansion coefficient maxima, and the line where the effective chemical potential is zero, $M=0$. By the symmetry of the functionals for the partition functions, the condition $M=0$ in fluids is analogous to the absence of an external field in the Ising model.

	\normalsize 
\section{Introduction}
A supercritical fluid is a state of matter that occurs when a substance is subjected to temperatures and pressures above its critical point (CP). Below CP the substance exhibits a liquid-gas coexistence. At CP, the distinction between the liquid and gas phases of a substance disappears, and beyond this point, the substance exists in a single continuous supercritical state. Despite the absence of first-order phase transitions above the critical point, fluids exhibit distinct liquid-like and gas-like behavior under supercritical conditions, and display a unique mix of properties. With densities comparable to liquids they possess low viscosity and high diffusivity like gases resulting in high solvent power. Moreover, the supercritical properties are instantly tunable by changing pressure and temperature. This is the reason why supercritical fluids have diverse applications across various industries~\cite{KMLetal14}.

Although the traditional liquid-gas phase distinction does not apply in the supercritical region, scientists continue to search for crossover regions or definitive lines that distinguish liquid-like and gas-like behaviors in the phase diagram. Naturally, such boundaries originate from CP and may either vanish away from it or extend infinitely far. Among known approaches to this problem are the Widom line, which represents the locus of correlation length maxima~\cite{FW69}; the Frenkel line~\cite{BGMetal17}, focusing on the dynamics of molecular motion; and the percolation line~\cite{CKS01}, related to spatial structures and marking the threshold where clusters of fluid or density fluctuations form a continuous, system-spanning network.

In the present work, we focus on the Widom line~\cite{XKBetal05,MS10,AKM17}, a concept based on thermodynamic response functions that represents a conditional boundary between liquid-like and gas-like regions in supercritical fluids. Thus, the Widom line \--- the locus of correlation length maxima emanating from CP~\cite{XKBetal05} \--- is treated as the continuation of the coexistence line above CP. Moreover, some thermodynamic response functions, such as isobaric heat capacity and isothermal compressibility, also exhibit maxima in the supercritical region. In the critical region beyond the critical point, the maxima lines of the correlation length and these response functions virtually merge into a single line~\cite{AKM17} and then diverge from each other as the temperature increases. Therefore, in the critical region, the Widom line is viewed as the locus of maxima of thermodynamic response functions~\cite{Proctor20}.

In our previous work~\cite{KD20}, the equation of state for a many-particle system interacting via a modified Morse potential was developed in the framework of a cell model. Recently~\cite{DKRP24,DKRP24arxiv}, the thermodynamic response functions, namely the isothermal compressibility, the thermal pressure coefficient, and the thermal expansion coefficient, have been calculated for the same model system. The isothermal compressibility and thermal expansion coefficient both develop a maximum in some region beyond critical point. On the one hand, such response function maxima, as mentioned above, are considered as good candidates for crossover lines between gas-like and liquid-like behaviors in the supercritical region. On the other hand, they are also considered as natural continuation of the coexistence line. Since our model allows for analytical expressions for many thermodynamic quantities~\cite{KD20,DKRP24}, it is straightforward to calculate and build such continuations. As a logical sequence of our previous works for the cell model with the modified Morse potential, in this paper we present results for three lines that can be cosidered as continuations of the coexistence line beyond the critical point in the pressure-temperature plane. The first one is the locus of the isothermal compressibility maxima. This line, as will be shown, terminates at some temperature above the critical one. The second line is the locus of the thermal expansion coefficient, which also terminates at some point. The third one is a line that is derived from the same condition that the first-order phase transition line below the critical point - the condition on the effective chemical potential $M=0$. This line can be extended into the supercritical region without any limits in temperature or pressure.  

In Section~\ref{sec:eos}, we recall the form of the modified Morse potential, present the explicit expression for the equation of state, and illustrate pressure isotherms at temperatures below and above the critical one. In Section~\ref{sec:widom}, we develop two alternatives of the Widom line, as the maxima locus of the isothermal compressibility and as the locus of maxima of the thermal expansion coefficient. Then we introduce the supercritical crossover line obtained directly from the equation of state of the cell fluid model applying a condition on the effective chemical potential that is the same as is used below CP to obtain the coexistence line.

\section{\label{sec:eos} The equation of state}

In the present work, we utilize the results obtained in~\cite{KD20,DKRP24arxiv,DKRP24}. In~\cite{KD20}, the phase space of collective variables was used to calculate the grand partition function of a cell fluid model in the zero-mode approximation, leading to explicit forms of the equation of state (EoS) in both pressure-temperature-chemical potential and pressure-temperature-density terms.
These EoSs are applicable across a wide range of temperatures below and above the CP. Furthermore, in~\cite{DKRP24arxiv,DKRP24}, we have used these equations as the basis for calculating and graphically representing the thermodynamic response functions \--- specifically, the isothermal compressibility, thermal pressure coefficient, and thermal expansion \--- of the cell fluid model with the modified Morse potential in the supercritical region.

The modified Morse potential used to describe the interaction between particles is given by 
\begin{eqnarray}
	\label{def:mod_morse}
	U(r) & = & \varepsilon C_{H} \left[A {\rm e}^{-n_0(r-R_0)/\alpha} \right.
	\nonumber\\
	&& + \left. {\rm e}^{-\gamma(r - R_0)/\alpha} -2{\rm e}^{-(r-R_0)/\alpha}\right],
\end{eqnarray}
where $R_0$ is the coordinate of the potential minimum, $\alpha$ is an effective range of interaction, $\gamma$ and $n_0$ are parameters of the model. Other two constants $C_{H}$ and $A$ are expressed via $\gamma$ and $n_0$ as follows
\begin{equation}
	C_{H} = \frac{n_0}{n_0 + \gamma - 2}, \quad A = \frac{2 - \gamma}{n_0},
\end{equation}
where $\varepsilon$ is the depth of the potential well at $r=R_0$. Applying $\gamma=2$ reduces $U(r)$ to the ordinary Morse potential~\cite{Morse29}. For a more detailed discussion of such modified Morse potential, see Sections~1 and~2 in~\cite{KD20}, and Section~2 in~\cite{DKRP24arxiv}.

We use $\varepsilon$ and $v$ as natural units for energy and volume. Therefore, the following reduced quantities are defined: the reduced density $\rho^*=\rho v$, where $\rho = \langle N \rangle /V$; reduced temperature $T^* = k_{\rm B}T/\varepsilon$; reduced pressure $P^* = Pv/\varepsilon$; reduced isothermal compressibility $\kappa^*_{T} = \varepsilon \kappa_T/v$; and the reduced thermal expansion coefficient $\alpha^*_{P} = \varepsilon \alpha_{P} / k_{\rm B}$. In above expressions $V$ stands for volume, $T$ the temperature, $k_{\rm B}$ the Boltzmann constant, $P$ the pressure, $\kappa_T$ the isothermal compressibility, $\alpha_{P}$ the thermal expansion coefficient, $\langle N \rangle$ is the average number of particles, with averaging over the grand canonical distribution.

The equation of state obtained in~\cite{KD20} and rewritten in reduced quantities in~\cite{DKRP24,DKRP24arxiv} reads
\begin{eqnarray}\label{eq:eosMT}
	P^* & = & (1 + \tau)T^*_c \bigg[ E_\mu(M, T) 
	\nonumber\\
	&& + M \bar \rho_0 + \frac{1}{2} d \bar \rho_0^2 - \frac{a_4}{24} \bar \rho_0^4
	\bigg].
\end{eqnarray}
Let us explicitly present the expressions entering the right-hand side of this equation.

First, the quantity $M$ depends linearly on the chemical potential
\begin{align}\label{chem_pot}
	&	M = \frac{\tilde\mu}{W(0)} + g_1 - \frac{g_3}{g_4} d - \frac{1}{6} \frac{g_3^3}{g_4^2}, \\
	&	\tilde\mu=\mu-\mu_0(1+\tau),
\end{align}
where $\mu_0$ is some positive constant, $\tau$ is the relative temperature $\tau = (T - T_c) / T_c$, $T_c$ is the critical temperature. We will call $M$ the effective chemical potential.

The quantity $W(0)$ is expressed via parameters of the potential~(\ref{def:mod_morse}) as follows
\begin{equation}
	W(0) = \Phi^{(r)}(0) \left[ B - 1 + \chi_0 + \tau (\chi_0 + A_\gamma) \right],
\end{equation}
where
\begin{align*} 
	& B = 2 \gamma^3 e^{(1-\gamma)R_0/\alpha},
	\nonumber \\
	& A_\gamma = A e^{(n_0-\gamma)R_0/\alpha} \left( \gamma / n_0\right)^3, 
\end{align*}
and $\Phi^{(r)}(0)$ is the Fourier transform of the repulsive part of the potential at $\abs{\vb k}=0$
\begin{equation*}
	\Phi^{(r)}(0) = \varepsilon C_H 8\pi {\rm e}^{\gamma R_0/\alpha} \left(\frac{\alpha}{\gamma R_0}\right)^3.
\end{equation*}

The parameter $\chi_0$ is used in~\cite{KD20} to single out a contribution in the Fourier transform of the potential that is treated as a reference system defined in the reciprocal space, and is selected as $\chi_0 = 0.07$~\cite{KD20}.

The coefficients $g_n$ are given by the formulas:
\begin{align}
	& g_0 = \ln T_0, \qquad g_1 = T_1/T_0, \qquad g_2 = T_2/T_0 - g_1^2,  
	\nonumber \\
	& g_3 = T_3/T_0 - g_1^3 - 3g_1 g_2, 
	\nonumber\\
	& g_4 = T_4/T_0 - g_1^4 - 6 g_1^2 g_2 - 4 g_1 g_3 - 3 g_2^2, 
\end{align}
where $T_n(p,\alpha^*)$ are the following special functions
\begin{equation}
	T_n(p,\alpha^*) = \sum_{m=0}^{\infty} \frac{(\alpha^*)^m}{m!} m^n {\rm e}^{-pm^2}.
\end{equation}
Here $\alpha^*=v e^{\beta_c\mu_0}$, where $\beta_c$ is the critical value of the inverse temperature $\beta = (k_{\rm B}T)^{-1}$, and the parameter $p$ has the form
\begin{equation}
	p = \frac{1}{2T^*_c} \frac{\Phi^{(r)}(0)}{\varepsilon} [\chi_0 + A_\gamma].
\end{equation} 
The critical temperature found in~\cite{KD20} is $T_c^* = 4.995.$

Since $p$ is independent of temperature, the coefficients $g_n$ are also independent of temperature. The numerical values for other coefficients used in this paper are the same as those in~\cite[eqs.~(5), (23), and~(24)]{KD20}:
\begin{eqnarray}
	\label{params}
	\chi_0 = 0.07, & \quad \gamma = 1.65, \nonumber\\
	n_0 = 1.521, & \quad R_0/\alpha = 2.9544, \nonumber\\
	\alpha^* = 5.0 & \quad p = 1.0.
\end{eqnarray}

The quantity $d$ entering equations~\eqref{eq:eosMT} and~\eqref{chem_pot} is a function of temperature
\begin{equation}
	\label{def:D0}
	d = g_2 - \frac{1}{2} \frac{g_3^2}{g_4} - \frac{1}{\beta W(0)}.
\end{equation}
The condition $d = 0$ defines the critical temperature~\cite{KD20}
\begin{equation}
	T^*_c = \left(g_2 - \frac{1}{2} \frac{g_3^2}{g_4} \right) (B - 1 + \chi_0) \frac{\Phi^{(r)}(0)}{\varepsilon}.
\end{equation}  
The function $E_\mu(M, T)$ from the equation \eqref{eq:eosMT} is given by
\begin{eqnarray}\label{eq:E_mu}
	E_\mu (M, T) & = & - \frac{\ln (2\pi \beta W(0))}{2 N_v}  
	\nonumber\\
	&& +  g_0 - \frac{\beta W(0)}{2} 
	\left(\frac{\tilde\mu}{W(0)} \right)^{2} 
	\nonumber\\
	&& - \frac{g_3}{g_4} {M} \! - \frac{g_3^2}{2 g_4^2}  d - \frac{1}{24} \frac{g_3^4}{g_4^3}. 
\end{eqnarray}
Here the quantity $N_v$ defines the number of cubic cells in volume $V$ for the initial model.
In the thermodynamic limit, $N_v \to \infty$, and thus, the first term can be neglected. The term $\tilde{\mu}/W(0)$ can be expressed in terms of $M$ using~\eqref{chem_pot}. The temperature and the inverse temperature can always be expressed in terms of the reduced temperature and a corresponding critical value:
\begin{equation*}
	T = T_c(1+\tau), \quad \beta = \beta_c (1 + \tau)^{-1}.
\end{equation*} 

The quantity $\bar{\rho}_0$ is a solution to the following cubic equation
\begin{equation}\label{eq:ro_M}
	M + d \bar\rho_0 + \frac{g_4}{6} \bar\rho_0^3 = 0.
\end{equation}
For any $\tau > 0$, the latter equation has one real root
\begin{equation}\label{eq:ro_MT}
	\bar \rho_0 = \left(- \frac{3 M}{g_4} + \sqrt{Q_t}\right)^{1/3} - \left(  \frac{3 M}{g_4} + \sqrt{Q_t} \right)^{1/3},
\end{equation}
where
\begin{equation}
	Q_t = \left(  \frac{2d}{g_4}\right)^3 + \left( -\frac{3 M}{g_4}\right)^2, \qquad g_4<0.
\end{equation}
Thus, $\bar{\rho}_0$ is a function of the temperature and the chemical potential.

\begin{figure}[t!]
	\centering
	\includegraphics[width=0.5\textwidth]{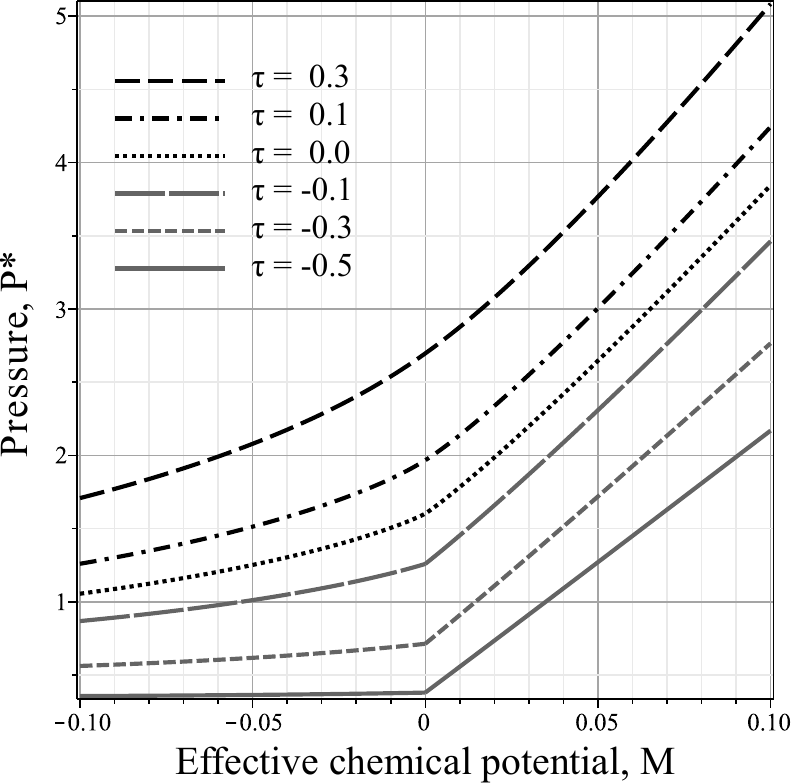} 
	\vskip-3mm\caption{Isotherms of the reduced pressure $P^*$ as a function of the effective chemical potential $M$. Black lines correspond to $T \geq T_c$. Grey lines correspond to $T < T_c$. 
	}\label{fig1}
\end{figure}
\begin{figure}[t!]
	\centering
	\includegraphics[width=0.5\textwidth]{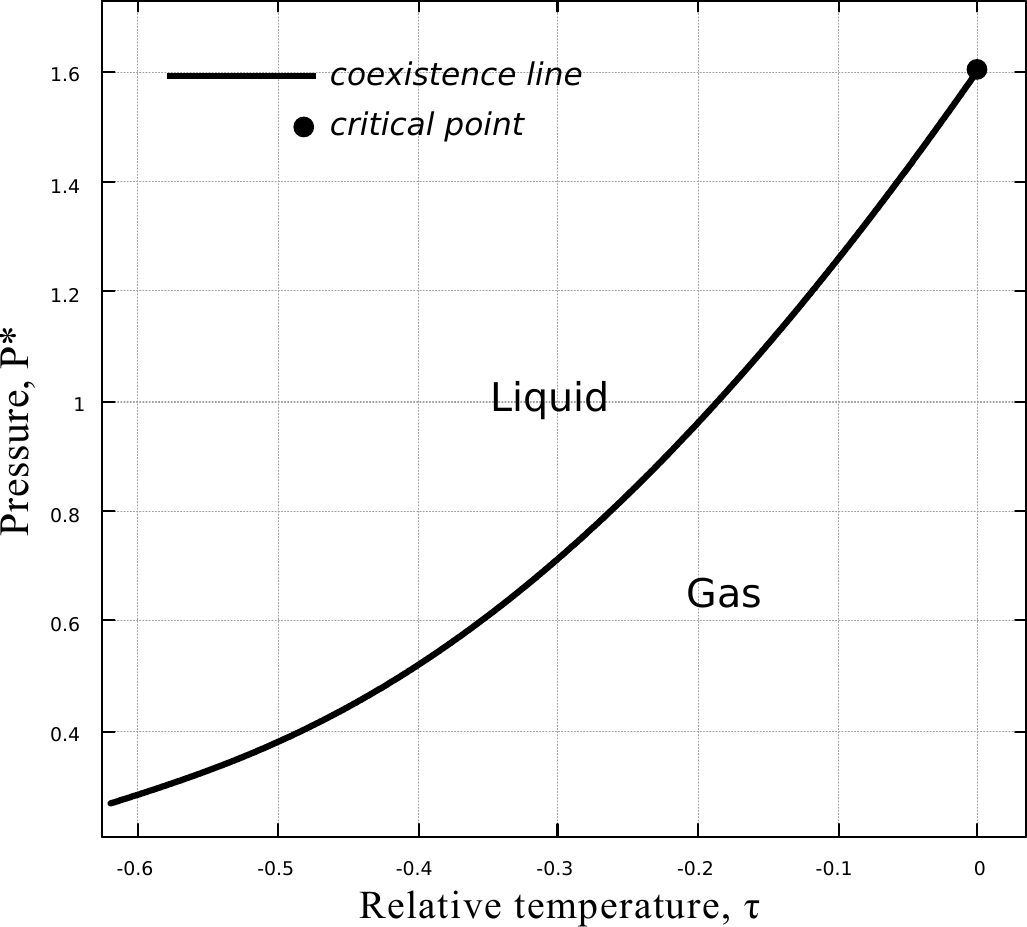} 
	\vskip-3mm\caption{Phase diagram in the $P^*-\tau$ plane for the cell fluid model with the interaction potential \eqref{def:mod_morse}. The result is based on the equation of state~\eqref{eq:eosMT}. 
	}\label{fig2}
\end{figure}

Figure~\ref{fig1} illustrates the relationship between the reduced pressure $P^*$ and the effective chemical potential $M$ for various relative temperature values $\tau$. Below the critical temperature ($\tau < 0$), each іsotherm contains a kink (three bottom lines in Figure~\ref{fig1}). This form indicates a first-order phase transition in the system. Within our approach~\cite{KD20} at $T<T_c$, negative values of the effective chemical potential $M$ imply the system is in a gaseous state. The condition $M=0$ is a point of phase coexistence. Positive values of the effective chemical potential correspond to the liquid state. In both cases, the pressure tends to increase with $M$, but the slope of the liquid isotherm is much steeper. At the critical temperature, the pressure is a smooth curve (black dotted line in Figure~\ref{fig1}) with an inflection point at $M=0$. Thus, at the critical point coordinates are defined by $M=0, \tau=0$, $P^*_c = 1.606$. Beyond the critical temperature, the pressure behaves as a monotonically increasing function, and no phase transitions are observed. Summing up, below the critical temperature, $M=0$ is the condition for phase coexistence. Substituting this condition into the EoS~\eqref{eq:eosMT}, we obtain a phase diagram in pressure-temperature plane (see Figure~\ref{fig2}). We will use the same condition to obtain a continuation of the coexistence line into the supercritical region. 

\section{\label{sec:widom} Crossover lines between gas-like and liquid-like behavior of the supercritical fluid.}

\begin{figure}[h!] 
	\centering \includegraphics[width=0.5\textwidth]{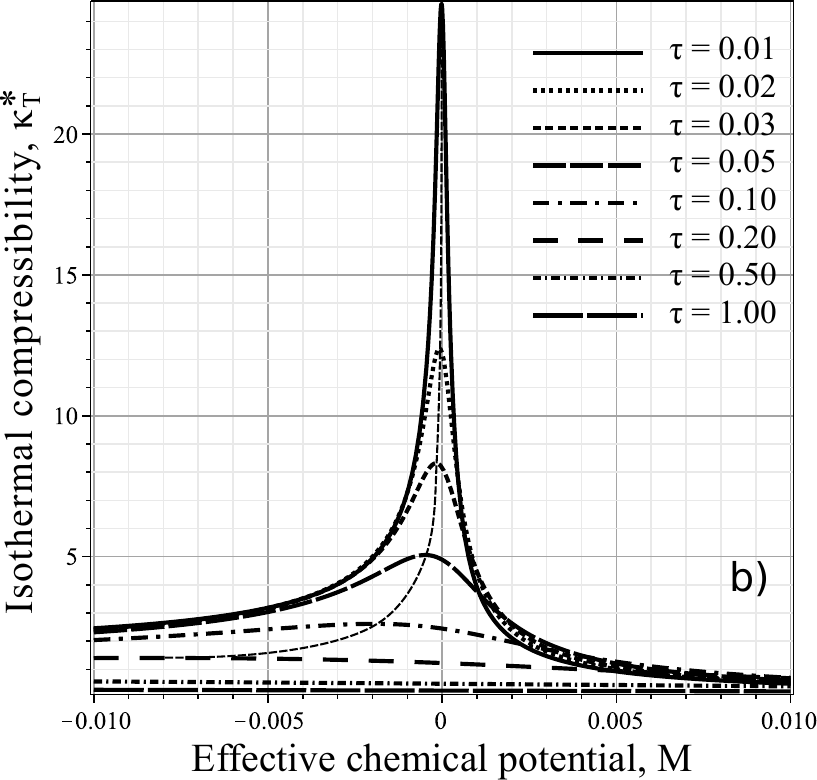}
	\vskip-3mm
	\caption{The reduced isothermal compressibility $\kappa^*_T$ as a function of the effective chemical potential $M$ for different temperatures $\tau = (T - T_c)/T_c$ at $T > T_c$. The plot focuses on a range of $M$ around its critical value $0$, thin dashed black line displays maxima locus of $\kappa^*_T$.
	}
	\label{fig3}
\end{figure}

In the supercritical region, substances no longer behave distinctly as either liquids or gases but exist as a single, continuous phase without a sharp liquid-gas boundary. However, fluids can still exhibit liquid-like and gas-like behaviors beyond the critical temperature. A way to represent a conditional boundary between these regions in supercritical fluids is the Widom line \--- a continuation of the coexistence line based on the locus of maxima of certain thermodynamic response functions~\cite{XKBetal05,MS10,SBGetal10}. This occurs because near CP, the system experiences significant fluctuations in density over increasingly large spatial scales. Although, the equation of state~\eqref{eq:eosMT} is derived in the mean-field-like approximation, we established in~\cite{DKRP24arxiv,DKRP24} that the isothermal compressibility of the cell fluid model diverges at CP and consequently displays maxima in the supercritical region (see Figures~\ref{fig3}). This suggests the possibility of deriving the Widom line and expanding the phase diagram in Figure~\ref{fig2} beyond CP.
In what follows we consider this in more detail.

The isothermal compressibility $\kappa_T$ is defined by
\begin{equation}
	\label{def:isotherm_compres}
	\kappa_T = -\frac{1}{V}\left(\frac{\partial V}{\partial P}\right)_{T, N}.
\end{equation}
In \cite{DKRP24arxiv,DKRP24} we derived expression for the reduced isothermal compressibility $\kappa^*_T \equiv \dfrac{\varepsilon \kappa_T}{v}$  in terms of thermodynamic derivatives suitable for the reduced form of the equation of state in pressure-temperture-chemical potential terms \eqref{eq:eosMT} 
It is expressed in terms of the reduced pressure $P^*$ and the reduced particle number density $\rho^*$ as follows
\begin{eqnarray}
	\label{eq:kappa_star_m1}
	\kappa^*_T & = & \frac{\varepsilon}{{\rho^*}^2} \left(\frac{\partial \rho^*}{\partial \mu}\right)_T
	\\
	\label{eq:kappa_star_m}
	& = & \frac{\varepsilon^2}{{\rho^*}^2} \left(\frac{\partial^2 P^*}{\partial \mu^2}\right)_T.
\end{eqnarray}
The reduced particle number density $\rho^*$ is found by 
\begin{equation}
	\rho^* \equiv \frac{\langle N \rangle}{V} v = \varepsilon\left(\frac{\partial P^*}{\partial \mu}\right)_{T,V}.
\end{equation}
Taking explicit derivatives, we arrive at
\begin{equation}\label{eq:density}
	\rho^* = \rho^*_c - M + \frac{ \bar \rho_0}{\beta W(0)}.
\end{equation}
The quantity $\rho^*_c$ in the equation \eqref{eq:density} is the critical density~\cite{KD20,DKRP24arxiv,DKRP24}
\begin{eqnarray}\label{eq:crit_dens}
	\rho^*_c & = &  g_1 - \frac{g_2 g_3}{g_4} + \frac{g_3^3}{3g_4^2}.
\end{eqnarray}
Its numerical value for parameters~\eqref{params} is $\rho^*_c = 0.978$.
Finally, the isothermal compressibility takes on the explicit form  
\begin{eqnarray}
	\kappa_T^* & = & \frac{\epsilon}{\rho^{*2} W(0)}
	\left\{   \frac{Q_t^{-1/2}}{\beta W(0) g_4} \right. \times
	\\ 
	&&\left. \times \left[ \bar \rho_0 - 2 \left( - \frac{3M}{g_4} + \sqrt{Q_t} \right)^{\frac{1}{3}} \right] -1 \right\}.
	\nonumber
\end{eqnarray}
Behavior of the reduced isothermal compressibility versus effective chemical potential at various fixed values of the reduced temperature is represented graphically in Figure~\ref{fig3} (narrow range around $M=0$). Right at the critical point the compressibility diverges. As a result, there is still a "memory" of the critical point, where $\kappa_T$ reaches local maxima. However, once the system moves further into the supercritical region, these maxima become less pronounced, and the compressibility decreases. 
Below the critical temperature, negative values of $M$ correspond to gaseous densities, while positive values of $M$ represent liquid densities. Above the critical point, the fluid becomes continuous. However, Figure~\ref{fig3} clearly shows significantly higher isothermal compressibility in the region where $M<0$, near the base of the maxima ridge, compared to the compressibility values in the $M>0$ region. In the first case ($M<0$), the supercritical fluid exhibits gas-like behavior. In the second case ($M>0$), it behaves like a liquid. Far beyond the critical point ($\tau > 0.5$), there is no noticeable difference in $\kappa_T$ values around the effective chemical potential $M=0$. At $\tau = 0.2, P^*>2.30$ the compressibility maxima become almost undistinguished, they completely vanish at $\tau > 0.28$. 
\begin{figure}[h!] 
	\centering \includegraphics[width=0.5\textwidth]{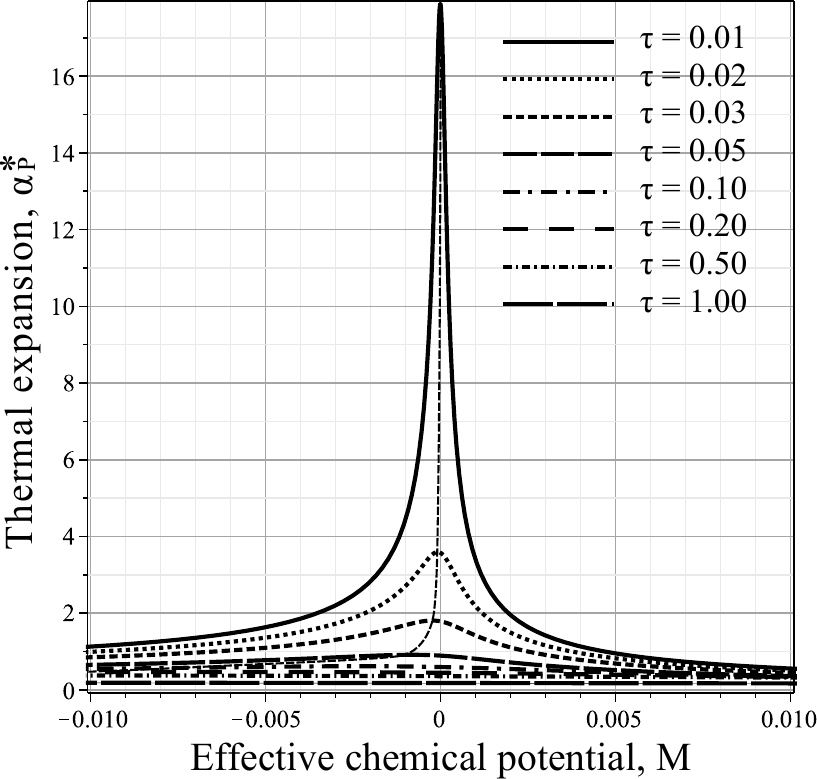}
	\vskip-3mm
	\caption{The reduced thermal expansion $\alpha^*_P$ as a function of the effective chemical potential $M$ for different temperatures $\tau = (T - T_c)/T_c$ at $T > T_c$. The plot focuses on a range of $M$ around its critical value $0$, thin dashed black line displays maxima locus of $\alpha^*_P$.
	}
	\label{fig3b}
\end{figure}
To identify the Widom line based on the locus of the isothermal compressibility maxima, we solve the equation
\begin{equation}\label{eq:maximize}
	\frac{\partial \kappa_T^*}{\partial M} = 0
\end{equation}
with respect to $M$.
Substituting the solution of \eqref{eq:maximize} into the equation of state \eqref{eq:eosMT}, one locates the Widom line in the pressure-temperature plane (see Figure~\ref{fig4}, dashed line), expanding the phase diagram displayed in Figure~\ref{fig2} beyond the critical point.

Similarly to the isothermal compressibility, similar arguments apply to the thermal expansion coefficient defined as
\begin{equation}
	\alpha_{P} = \frac{1}{V} \left(\frac{\partial V}{\partial T}\right)_{P,N}.
\end{equation}
Its reduced counterpart is calculated based on EoS~\eqref{eq:eosMT} by formula
\begin{eqnarray}
	\alpha^*_P & = & \frac{1}{T^*_c \rho^*}
	\left[ 
	-\left(\frac{\partial \rho^*}{\partial \tau}\right)_{\mu}
	\right. \nonumber\\
	&& + \left.\left(\frac{\partial \rho^*}{\partial \mu}\right)_{T}
	\left(\frac{\partial P^*}{\partial \tau}\right)_{\mu}
	\left(\frac{\partial P^*}{\partial \mu}\right)^{-1}_{T} 
	\right].
\end{eqnarray}
Figure~\ref{fig3b} illustrates the dependence of $\alpha^*_{P}$ on $M$ at some given values of $\tau$, at $T>T_c$. The thermal expansion coefficient also develops a ridge of maxima in the supercritical phase, that almost vanishes at $\tau > 0.35, P^*>2.75$.
To identify the Widom line based on the locus of the thermal expansion coefficient maxima, we solve the equation
\begin{equation}\label{eq:maximize_alpha}
	\frac{\partial \alpha_P^*}{\partial M} = 0
\end{equation}
with respect to $M$.
Substituting the solution of \eqref{eq:maximize_alpha} into the equation of state \eqref{eq:eosMT}, one locates another Widom line in the pressure-temperature plane (see Figure~\ref{fig4}, dash-dotted line). This is the second option for the separation line between liquid-like and gas-like fluid.

The third choice for the boundary between liquid-like and gas-like regions beyond CP is based on the condition $M=0$. Within the present approach, the condition $M=0$ holds along phase coexistence up to the critical point. Thus, the implementation of the condition $M=0$ above CP seems to be a natural continuation of the coexistence line into the supercritical region. On the other hand, taking into account the functional form of the grand partition function~\cite{KD16,KR09}, the condition of $M=0$ is equivalent to the absence of an external field in the Ising model. This analogy is interesting from the perspective of direct mapping between the Ising model and fluids~\cite{Kulinskii10jcp,BK11}. In Fig.~\ref{fig4} the separation line based on this condition $M=0$ is given by the dotted curve. In the critical region ($\tau<0.02$), beyond CP, the maxima locus of the response functions and the $M=0$ based line visually merge into a single line.

\begin{figure}[h!]
	\centering \includegraphics[width=0.5\textwidth]{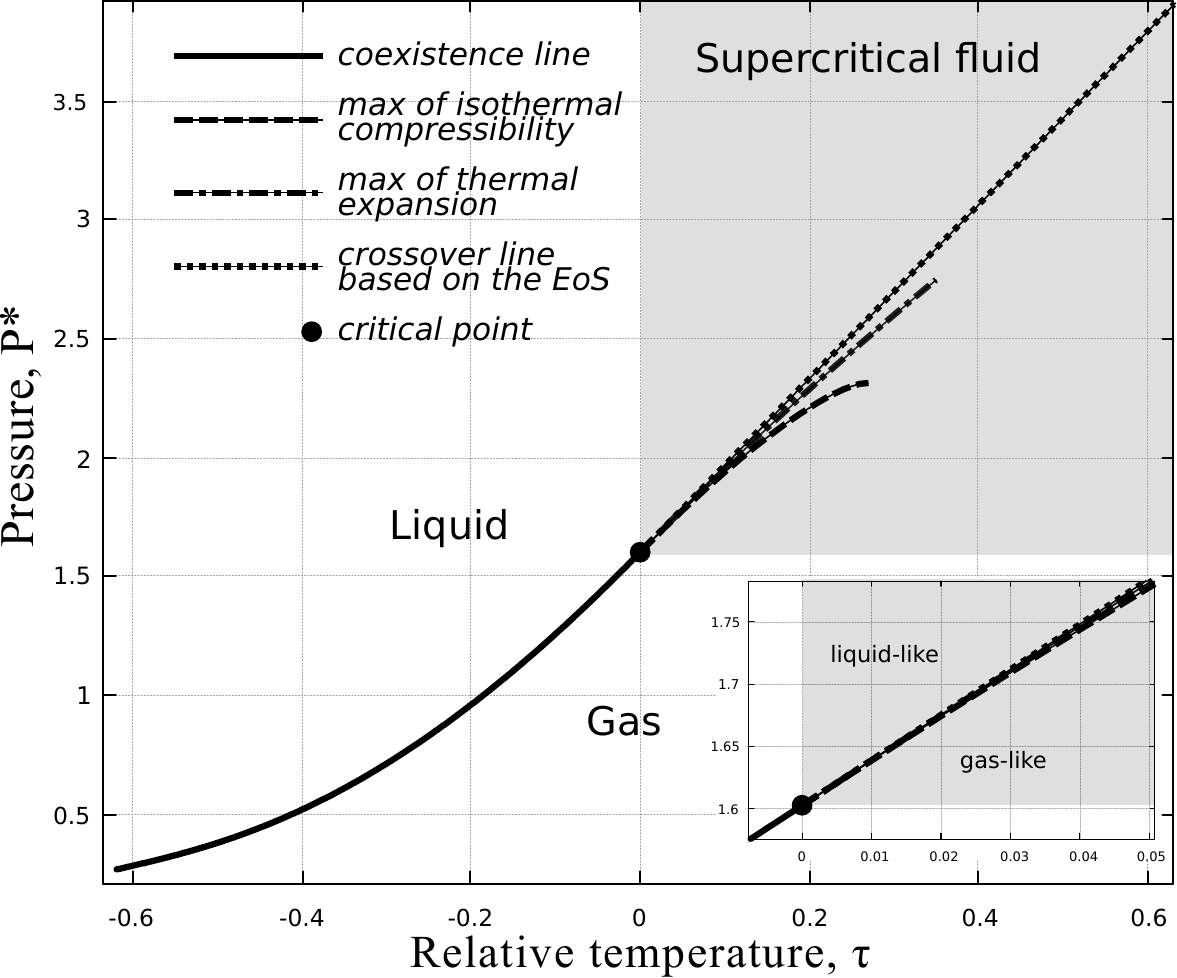}
	\vskip-3mm\caption{Complete pressure-temperature phase diagram of the cell fluid model with the interaction potential \eqref{def:mod_morse}. Comparison of the Widom line (maxima locus of isothermal compressibility and thermal pressure) and the crossover between gas-like and liquid-like structures of the supercritical cell fluid based on combination of the equation of state~\eqref{eq:eosMT} and the condition $M = 0$. An inset in the right bottom corner focuses on a range of $\tau$ around its critical value $-0.01 < \tau < 0.05$.	}\label{fig4}
\end{figure}

\section{Conclusions}
The supercritical region of the phase diagram is studied for the cell fluid model with the modified Morse potential. Three lines are found that can be considered as good candidates for a separation boundary between gas-like and liquid-like regions in the supercritical phase. The first line is the locus of the isothermal compressibility maxima. The second line is the locus of the thermal expansion coefficient maxima. The third line is the location of states where the condition $M=0$ holds true. This third option seems to be a natural extension of the coexistence line beyond the critical point, since the same condition is used both below and above $T_c$. A noticeable difference between two loci of thermodynamic response functions maxima and the line obtained from $M=0$ is that the latter one is extended over the phase diagram without any limits, while the first two have their termination points. Investigation of how physical quantities change their behavior when crossing these lines will be the subject of our future studies.

\vskip3mm \textit{This work was supported by the National Research Foundation of Ukraine under the project No.~2023.03/0201.}

\end{document}